\documentclass[11pt]{article}

\usepackage[utf8]{inputenc}
\usepackage[T1]{fontenc}
\usepackage{lmodern}
\usepackage[margin=1in]{geometry}
\usepackage{graphicx}
\usepackage{booktabs}
\usepackage{array}
\usepackage{amsmath,amssymb}
\usepackage{microtype}
\usepackage[font=small,labelfont=bf]{caption}
\usepackage[numbers,sort&compress,square]{natbib}
\usepackage{xcolor}
\usepackage[hidelinks]{hyperref}

\hypersetup{
  pdftitle={Concurrency Response of Plain Global Loads on the NVIDIA H100},
  pdfauthor={Somashekar Manjunath, Rahul Ramachandra M},
  pdfsubject={GPU memory-concurrency microbenchmarking}
}

\graphicspath{{figures/}}

\newcommand{\pct}{\%}
\title{\bfseries Concurrency Response of Plain Global Loads\\ on the NVIDIA H100}

\author{
  Somashekar Manjunath \\ CoreOptX Technologies \\ \texttt{somashekar@coreoptx.com}
  \and
  Rahul Ramachandra M \\ CoreOptX Technologies \\ \texttt{rahul.m@coreoptx.com}
}

\date{}

\begin{document}
\maketitle

\begin{abstract}
\noindent The bandwidth a memory-bound GPU kernel sustains is set by how many bytes it keeps in flight. We use Little's Law here as throughput accounting, not as a measured hardware pool. CUDA fills that budget on Hopper through plain loads (\texttt{ld.global}) and asynchronous copies (\texttt{cp.async}), among other paths; we characterize their concurrency response with clean-room microbenchmarks on three H100 SXM5 dies.

Our main result concerns the plain-load path: attained LDG bandwidth peaks at a small offered per-thread load ($K \approx 2$) and then \emph{declines}, by about 35\pct{} from $K{=}2$ to $K{=}8$ at our primary configuration. The decline survives a fixed-work control matching total \emph{issued logical loads} across $K$, ascending and reversed sweep orders, and replication on two dies with the same instrument ($-35.0\pct$ and $-35.2\pct$). Separately profiled counters show DRAM bytes nearly constant over $K{=}2\!\to\!8$ while L2-sector traffic rises, and a $40\times$ nominal allocation-size sweep (512~MB to 20~GB, all above the ${\sim}50$~MB L2; no address trace) leaves the decline essentially unchanged, disfavoring a simple allocation-size dependence. Because the L2 hit-rate nonetheless rises with $K$ at every allocation, the aggregate request stream does change with $K$; we report $K$ as \emph{offered software ILP} and leave the hardware mechanism open.

A preliminary survey adds a matched \texttt{cp.async}-versus-plain-load comparison ($2.1$--$2.9\times$ at high offered depth, two dies), a die-B same-CTA two-stream observation whose companion die-C check differs and is not pooled, and a cross-die primitive baseline.
\end{abstract}

\begin{quote}\small
\textbf{Scope and evidence.} The plain-load (LDG) result in Section~\ref{sec:decline} is fixed-work-controlled, replicated across two dies, cross-checked with separately profiled hardware counters, and includes a $40\times$ allocation-size sweep (nominal, not address-traced) that disfavors a simple allocation-size dependence. The \texttt{cp.async} comparison is a two-die, timing-only survey. The two-stream result is a die-B, single-session timing observation; its companion die-C exploratory check differs and is reported rather than pooled. Every claim is operational (what the hardware delivers for a given offered load), and we do not infer a physical queue, miss-handling structure, or pool. Numbers are our own microbenchmark measurements or cited prior work, never mixed.
\end{quote}

\section{Introduction}
\label{sec:intro}
For a memory-bound loop, Little's Law fixes attained bandwidth as bytes-in-flight divided by round-trip latency, an accounting identity among measured quantities, not a claim about any particular hardware buffer. On an H100 SXM5, unloaded global latency is a few hundred nanoseconds and peak HBM3 bandwidth is several terabytes per second, so a kernel must keep on the order of ten kilobytes per SM in flight to saturate the memory system. That concurrency comes from thread-level parallelism (more warps) or instruction-level parallelism (more independent in-flight loads per thread), and, on Hopper, from distinct instructions: plain loads with per-warp scoreboard completion and \texttt{cp.async} copies that write shared memory with no destination register.

Offered-request behavior was characterized on early GPUs. Wong et al. established the microbenchmarking methodology on GT200~\citep{wong2010}; outstanding-miss limits were later inferred for Fermi- and Kepler-generation parts~\citep{nugteren2014,lashgar2016}. From Volta onward the L1 is a streaming design; Volta was dissected in detail by microbenchmark~\citep{jia2018}, its memory system modeled and validated on Titan~V~\citep{khairy2018}, and the memory hierarchy modeled by microbenchmark~\citep{mei2017}, and Hopper-generation studies followed~\citep{hanindhito2024,luo2025,tawa2026}. Against that background we do not claim to be first to measure any of these units. We contribute a controlled measurement of how plain-load bandwidth responds to offered per-thread load on Hopper's streaming L1, replicated across dies, plus a die-C $40\times$ nominal allocation-size sweep that disfavors a simple allocation-size dependence, and a preliminary per-mechanism survey.

This paper reports:
\begin{enumerate}
\item A fixed-work-controlled \textbf{post-peak decline} in plain-load bandwidth: attained LDG bandwidth peaks near $K{=}2$ and falls as offered loads per thread rise, at matched issued logical work (mismatch ${\le}0.0246\pct$; separately profiled, near-constant DRAM traffic), on two dies with the same instrument (Section~\ref{sec:decline}).
\item That the endpoint decline is essentially unchanged across a $40\times$ nominal allocation-size range (512~MB $\to$ 20~GB) on die-C, disfavoring a simple allocation-size dependence (Section~\ref{sec:decline}).
\item A matched comparison in which \texttt{cp.async} sustains $2.1$--$2.9\times$ the bandwidth of an equivalent plain-load kernel at high offered depth, on two dies (Section~\ref{sec:cpasync}).
\item A die-B same-CTA two-stream displacement measurement; its companion die-C exploratory check differs and is not pooled (Section~\ref{sec:twostream}).
\item A primitive baseline (latency ladder and streaming ceiling) consistent across three dies (Section~\ref{sec:baseline}).
\end{enumerate}

Only the first two are fixed-work-controlled and counter-checked; \texttt{cp.async} is a cross-die timing-only comparison, while the two-stream observation is die-B session-specific with a non-pooled die-C check. We treat all survey items as preliminary. Section~\ref{sec:concl} lists the confirmatory work each result needs.

\section{Background}
\label{sec:bg}
\paragraph{Little's Law on GPUs.} For a memory-bound loop, attained bandwidth equals bytes in flight divided by latency, and either more warps or more independent in-flight loads per thread can raise the numerator~\citep{volkov2016}. This is throughput-accounting intuition, a relation among measured quantities, not evidence for a measured hardware budget or pool. At fixed occupancy, per-thread ILP (our axis $K$) is the parameter that varies offered in-flight bytes.

\paragraph{Prior supply measurements.} Nugteren et al. inferred a per-warp outstanding-miss limit of six and a per-SM limit near sixty-four on Fermi~\citep{nugteren2014}; Lashgar et al. inferred at least about 45 pending-table-like entries and on the order of 1{,}408 unique outstanding requests in flight for their Kepler configuration~\citep{lashgar2016}.

\paragraph{Streaming L1 and Hopper.} From Volta the L1/shared-memory complex is a streaming design; Ampere introduced the \texttt{cp.async} asynchronous-copy path, and Hopper adds the TMA, bulk-asynchronous copies, and a physically partitioned ${\sim}50$~MB L2. Existing Hopper studies characterize several of these features~\citep{hanindhito2024,luo2025,tawa2026}; we add a controlled plain-load response curve, a die-C $40\times$ nominal allocation-size sweep, and a preliminary per-mechanism survey.

\section{Methodology}
\label{sec:method}
\paragraph{Instruments.} Each kernel is implemented from public ISA and architecture documents. The plain-load kernel is SASS-audited to issue exactly $K$ \texttt{ld.global} instructions per iteration with zero register spills at every $K$; loaded values are XOR-accumulated into a sink the compiler cannot prove dead but that never stores on the timed path, so loads cannot be eliminated. Latency probes chase a dependent pointer chain, which precludes elimination by construction.

\paragraph{The axis $K$.} $K$ is the number of independent \texttt{ld.global} instructions the source issues per thread per iteration before consuming them, that is, the offered software ILP. We do not measure how many are simultaneously in flight in hardware, and treat every statement about $K$ as operational.

\paragraph{Fixed-work control.} A fixed-iteration $K$ sweep moves total traffic proportional to $K$ and is naturally run ascending, confounding offered load with total work and with run order. We break this by setting each cell's iteration count to $L/(\text{threads}\,{\times}\,K)$ for a fixed target load count $L$, which matches total issued logical loads across $K$ (${\le}0.0246\pct$; physical DRAM traffic is audited separately with counters), and by sweeping $K$ both ascending and reversed. The warp's access is contiguous and fully coalesced at every $K$; only the per-thread offered count changes (SASS-verified).

\paragraph{Allocation-size sweep (die-C).} To test for an allocation-size dependence, we repeat the fixed-work sweep at HBM buffer allocations of 512~MB, 2~GB, 8~GB, and 20~GB, spanning from about ten times the ${\sim}50$~MB L2 to $400\times$ it, and read the decline and the L2 hit-rate at each. The issued read volume is 16{,}785~MiB at every size, so the issued-volume-to-allocation ratio varies from ${\approx}33{:}1$ (512~MiB) to ${\approx}0.84{:}1$ (20{,}000~MiB, bounding unique coverage at ${\le}84\pct$). No address trace is collected, so per-line reuse is not observed: this is a nominal allocation-size sweep, not a verified unique-working-set sweep. The measured DRAM traffic is constant (${\sim}17.6$~GB) at all four sizes.

\paragraph{Hardware counters.} On the primary cells we capture, with \texttt{ncu}, the DRAM bytes moved, the issued global-load sectors, the L2 sectors, and the L2 hit-rate; the exact counter identifiers are documented in our analysis scripts. The parser excludes only the short warm-up launch by rule and reports the median with range over the remaining full-size launches. Under the profiler the effective clock is depressed (about 1.6 vs 1.98~GHz native), so we use these counters to check \emph{how many bytes and sectors moved}, not for timing; the timing comes from the separate un-profiled run.

\paragraph{Environment.} All dies are H100 80~GB HBM3 SXM5 (132 SMs, ${\sim}50$~MB L2, 3{,}350~GB/s datasheet). The die-C controlled runs used driver 595.71.05 and CUDA 12.9 at a steady 1{,}830~MHz SM clock (700~W limit, ECC on, persistence on, no throttling flags during collection); the full environment capture is retained with the die-C raw data. The die-B session's environment is documented only to the extent of its session-provenance summary (its raw environment capture is not included).

\paragraph{Timing, residency, statistics.} Intra-SM intervals use \texttt{clock64}; cross-SM windows use a runtime-calibrated \texttt{\%globaltimer}, and we never difference \texttt{clock64} across SMs. Bandwidth is bytes per calibrated nanosecond; any systematic calibration error is shared within a session and cancels in the decline percentages, which are ratios of bandwidths under one calibration, while absolute bandwidths carry it. For HBM-targeted cells, a per-cell check flags any attained bandwidth above the 3{,}350~GB/s datasheet (impossible for true DRAM traffic); the fixed-work HBM sessions have no such violations (a residency screen, not a proof; L2-targeted cells legitimately exceed it by design). The unit is the kernel launch ($n{=}3$ per arm for the die-B controls; $n{=}10$ for the die-C primary; $n{=}1$ per allocation $\times$ order cell in the die-C sweep); knees use a frozen 95\pct-plateau rule with a stable three-cell tail (fixed before the die-B/die-C sessions; the dated preregistration record is retained in the program archive and is not included in this submission).
\section{Results}
\label{sec:results}

\subsection{Primitive baseline}
\label{sec:baseline}
\begin{figure}[tbp]\centering
\includegraphics[width=0.82\linewidth]{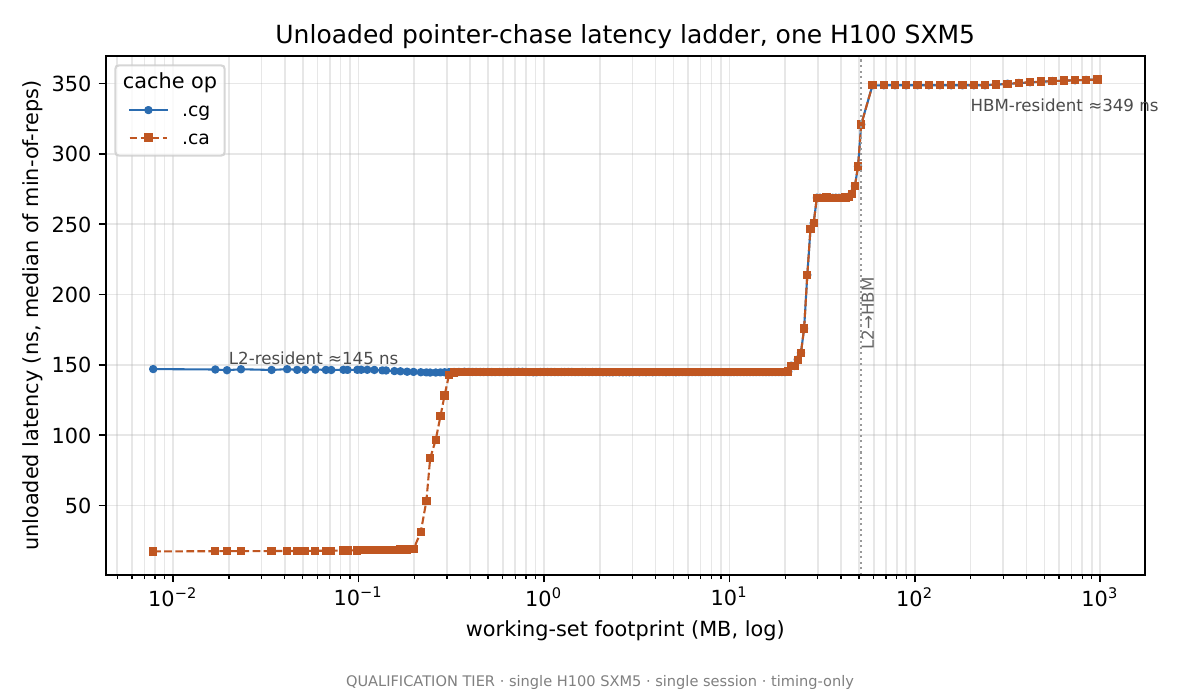}
\caption{Unloaded pointer-chase latency versus working-set footprint (verified by construction: the chase touches every element of its buffer), \texttt{.ca} and \texttt{.cg}, single SM: L1 ${\sim}17$~ns, L2 ${\sim}145$~ns, intermediate ${\sim}269$~ns (29--44~MB), HBM ${\sim}353$~ns.}
\label{fig:ladder}
\end{figure}

\begin{figure}[tbp]\centering
\includegraphics[width=0.82\linewidth]{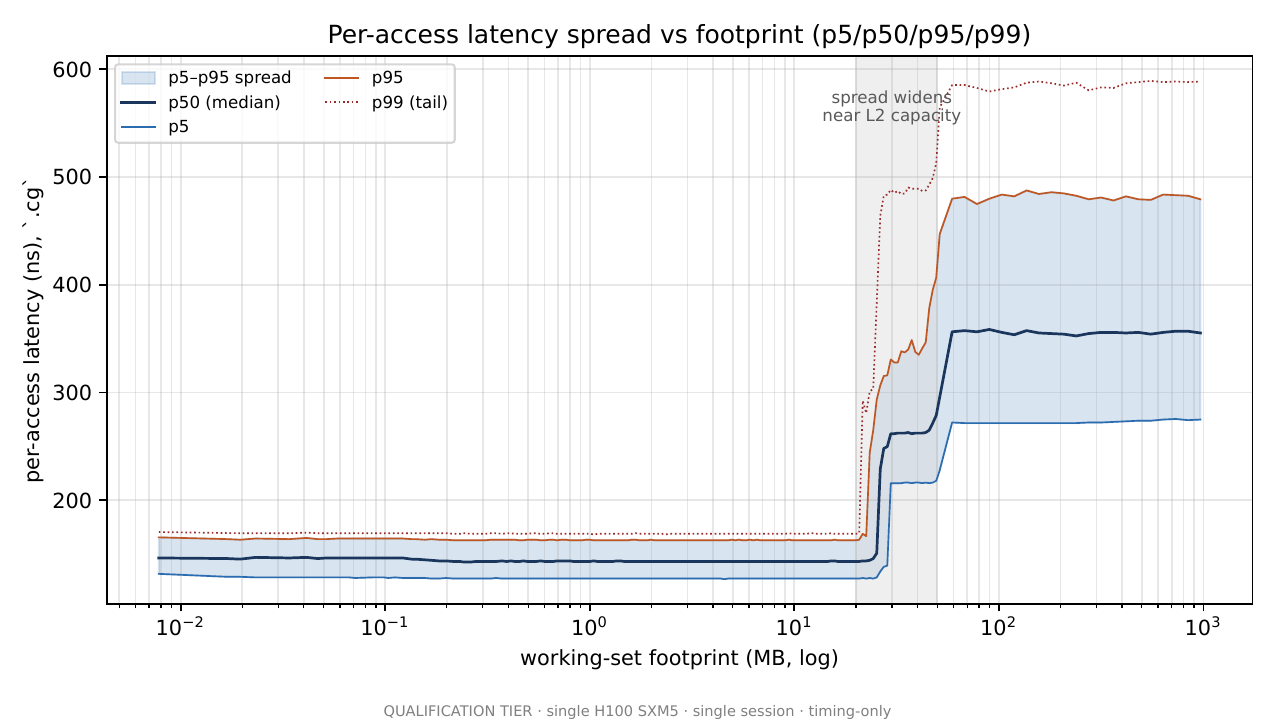}
\caption{Per-access latency spread (p5/p50/p95/p99) versus footprint; the spread widens near L2 capacity.}
\label{fig:dist}
\end{figure}

Unloaded pointer-chase latency on a single SM resolves a four-level ladder (Fig.~\ref{fig:ladder}): an L1 hit at ${\sim}17$~ns (\texttt{.ca}, 8~KB); an L2-resident plateau at ${\sim}145$~ns, flat from 8~KB to about 20~MB; an intermediate plateau at ${\sim}269$~ns between roughly 29 and 44~MB, as the working set approaches the ${\sim}50$~MB L2; and an HBM-resident plateau at ${\sim}353$~ns above 59~MB. The \texttt{.cg} operator bypasses the L1 (147~ns at 8~KB against 17~ns for \texttt{.ca}), which validates the probe. The per-access latency spread (Fig.~\ref{fig:dist}) widens near L2 capacity; we report this as a spread, not a resolved near/far slice geometry, which single-SM data cannot establish. Streaming read bandwidth is a median 91--94\pct{} of the 3{,}350~GB/s datasheet across dies; the die-C figure (median 91.2\pct; best-of-31 launches 91.3\pct) is the K1 acceptance-gate row used by the suite (read, 4~B width, ILP~1). In the raw logs the recorded per-cell figure is the best of the $N$ launches (the field naming is documented with the raw data). Die-A and die-C agree to within a few percent on every primitive checkable from their raw data, including the intermediate 269~ns plateau on both; die-B agreement is summarized from its session records (die-B raw data not included).

\subsection{A controlled post-peak decline with an allocation-size sweep}
\label{sec:decline}
\begin{figure}[tbp]\centering
\includegraphics[width=\linewidth]{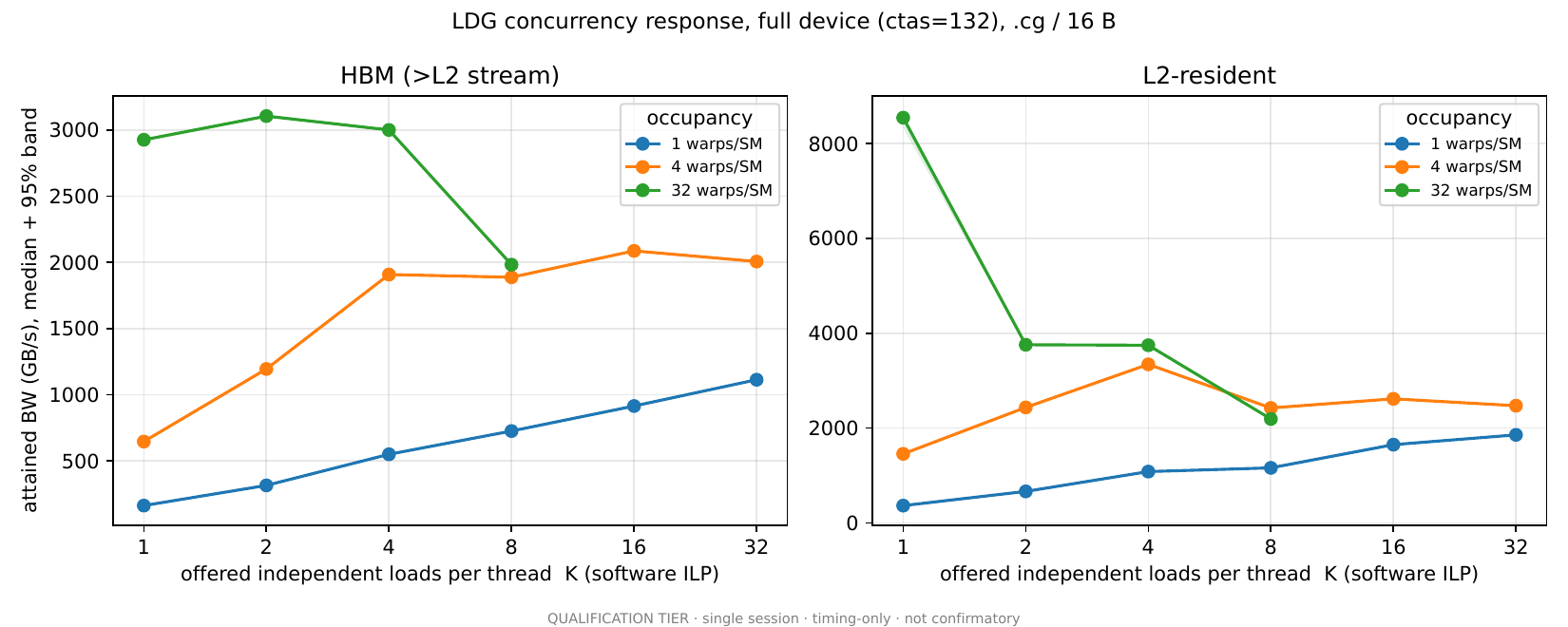}
\caption{Plain-load response curves: attained bandwidth versus $K$ at 132 CTAs, three occupancies, \texttt{.cg}~/~16~B, HBM and L2 backends; median with a 2.5--97.5 percentile band. Qualification-tier, single die-A session (fixed-iteration sweep).}
\label{fig:curves}
\end{figure}

\begin{figure}[tbp]\centering
\includegraphics[width=0.82\linewidth]{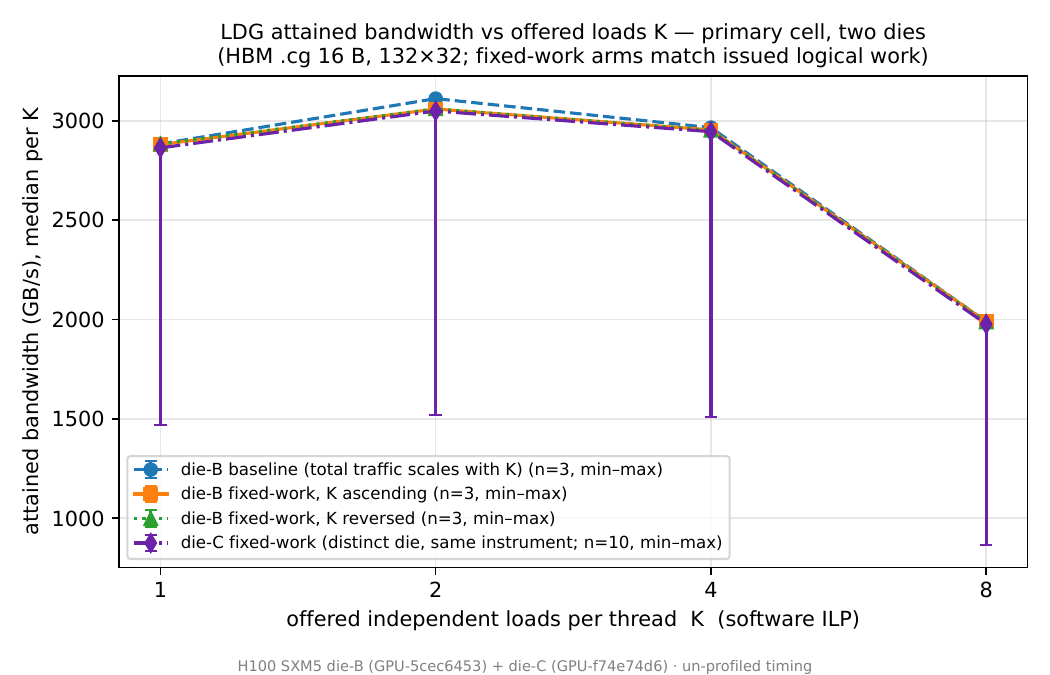}
\caption{Primary cell: die-B baseline (traffic grows with $K$) versus fixed-work ascending and reversed ($n{=}3$, min--max whiskers), and die-C fixed-work ($n{=}10$, min--max whiskers). The decline persists at matched issued logical work (mismatch ${\le}0.0246\pct$) on both dies. The long lower whiskers on the die-C series are the two disclosed anomalous launches (Section~\ref{sec:decline}); medians are insensitive to them. Companion: the die-C allocation-size sweep (512~MB $\to$ 20~GB) with the decline flat and the L2 hit-rate at $K{=}8$ unchanged (${\sim}56\pct$).}
\label{fig:oversub}
\end{figure}

Attained LDG bandwidth peaks near $K{=}2$ and then falls as $K$ rises (Fig.~\ref{fig:curves}, Fig.~\ref{fig:oversub}). At the primary cell (HBM, \texttt{.cg}, 16~B, 132 CTAs $\times$ 32 warps):

\begin{table}[htbp]
\centering\small
\begin{tabular}{@{}l l r r r@{}}
\toprule
Arm & Issued logical work & $K{=}2$ & $K{=}8$ & change \\
\midrule
die-B baseline (fixed iterations)   & grows with $K$          & 3111.0 GB/s & 1989.5 GB/s & $-36.1\pct$ \\
die-B fixed-work, ascending ($n{=}3$) & matched (max 0.0246\pct) & 3058.8      & 1988.7      & $-35.0\pct$ \\
die-B fixed-work, reversed ($n{=}3$)  & matched (max 0.0246\pct) & 3059.2      & 1987.0      & $-35.0\pct$ \\
die-C fixed-work, ascending ($n{=}10$) & matched (max 0.0246\pct) & 3048.6      & 1976.7      & $-35.2\pct$ \\
\bottomrule
\end{tabular}
\end{table}

The fixed-work arms match total issued \emph{logical} loads across $K$ (${\le}0.0246\pct$), so the decline is not an artifact of the larger workload a fixed-iteration sweep runs at high $K$; the ascending and reversed sweeps agree, so it is not a simple run-order trend; and it reproduces on two distinct dies, measured with the same instrument, at $-35.0\pct$ (both orders) and $-35.2\pct$. These are two device replications using the same instrument, not two independent methods. The controlled data span 16 and 32 warps at 132 CTAs. A separate die-A qualification grid includes 66- and 132-CTA cells and reproduces a post-peak shape, but it lacks the full fixed-work/order/counter/allocation-sweep design and is not part of the controlled inference.

\paragraph{Exact-work accounting and L2-targeted control.} ``Matched'' in the table uses the exact product of iteration count and per-thread offered count over the full $K{=}\{1,2,4,8\}$ sweep; its maximum mismatch is $0.024576\pct$. A coarser whole-MiB volume field in the raw logs is rounded and is not used for that bound. On the separately collected die-B L2-targeted backend at 132 CTAs $\times$ 32 warps, the same fixed-work construction falls $41.19\pct$ ascending and $41.22\pct$ reversed ($n{=}3$/arm). This is additional fixed-work timing evidence, not part of the HBM allocation-sweep or counter analysis.

\paragraph{Allocation-size sweep (die-C, $n{=}1$ per cell).} Repeating the fixed-work sweep at HBM buffer allocations from 512~MB to 20~GB, a $40\times$ nominal range, all above the ${\sim}50$~MB L2, leaves the endpoint decline essentially unchanged. Each allocation $\times$ order cell below is a single sweep (the $n{=}10$ repetition is the primary cell above); across all eight cells the decline spans $34.95$--$35.30\pct$:

\begin{table}[htbp]
\centering\small
\begin{tabular}{@{}l r r r r@{}}
\toprule
allocation & decline (asc) & decline (rev) & L2 hit @$K{=}8$ & L2 hit @$K{=}2$ \\
\midrule
512 MB & 35.20\pct & 35.30\pct & 56.5\pct & 0.4\pct \\
2 GB   & 35.18\pct & 34.95\pct & 55.9\pct & 0.8\pct \\
8 GB   & 35.16\pct & 35.07\pct & 55.8\pct & 0.9\pct \\
20 GB  & 35.08\pct & 35.10\pct & 55.7\pct & 0.9\pct \\
\bottomrule
\end{tabular}
\end{table}

\begin{figure}[tbp]\centering
\includegraphics[width=0.82\linewidth]{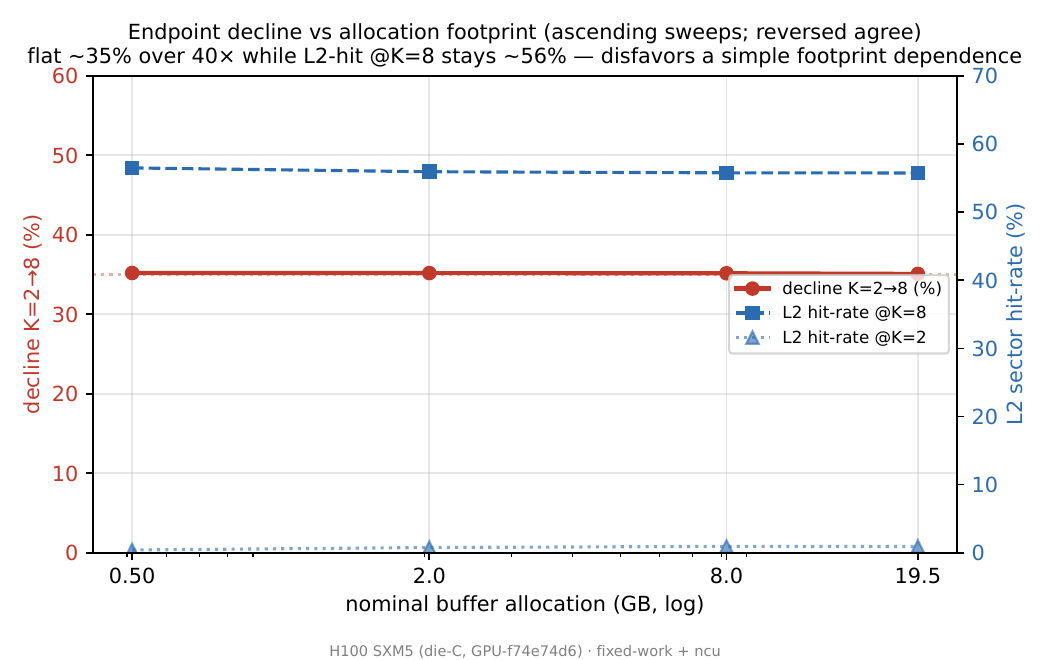}
\caption{Decline ($K{=}2\!\to\!8$) and L2 hit-rate (from the ascending-order NCU profiles) versus nominal buffer allocation over $40\times$ (die-C; issued volume constant at 17.6~GB per sweep); decline flat at ${\sim}35\pct$ while L2 hit-rate at $K{=}8$ stays ${\sim}56\pct$, disfavoring a simple allocation-size dependence.}
\label{fig:footprint}
\end{figure}

This \textbf{disfavors a simple allocation-size dependence} (Fig.~\ref{fig:footprint}): all tested allocations already exceed the L2; the decline does not attenuate as the allocation grows to $400\times$ the cache, even as the issued-volume-to-allocation ratio falls from ${\approx}33{:}1$ to ${\approx}0.84{:}1$. It does not by itself exclude cache-path effects: the L2 hit-rate rises with $K$ (from under 1\pct{} to ${\sim}56\pct$) at every allocation size, so the aggregate request stream, its interleaving and short-range reuse, changes with $K$ even though each warp's accesses remain contiguous and coalesced. Aggregate-interleaving and L2-path explanations therefore remain open, and a discriminating experiment (varying the $K$-dependent address ordering at fixed coalescing, occupancy, and logical bytes, with allocations below and around the L2, plus address-trace coverage verification) is left as future work.

\paragraph{Counter audit (separately profiled).} Over $K{=}2\!\to\!8$ the measured DRAM bytes are nearly constant: $-0.76\pct$ at the die-B primary, $-0.01\pct$ at the die-C 20~GB cell, and $-1.57\pct$ at the die-C 512~MB cell (the largest decline among audited HBM-allocation cells). Across the 14 complete E3 profiles containing $K{=}2$ and $K{=}8$, the DRAM change spans $-1.57\pct$ to $+1.25\pct$. Six incomplete E3 profiles (three HBM and three L2) contain \texttt{LaunchFailed} and are excluded from this cross-cell range. In the die-B primary and die-C HBM-allocation series, issued load sectors are matched while un-profiled attained bandwidth falls by about a third and L2-sector traffic rises about 50\pct. (In cross-occupancy audit cells the issued-sector totals scale with thread count, as expected; the matched-sector statement is specific to the primary and HBM-allocation series.) The audited quantities exclude the tested work-volume and simple sweep-order explanations; they do not identify the mechanism.

\paragraph{Repeatability and two anomalies.} Across $n{=}10$ launches of the primary cell, eight agree to within 0.6\pct{} at every $K$ (0.2\pct{} at $K{=}2$), with per-launch paired declines of $34.9$--$35.2\pct$. Launch~9 ran 0--8\pct{} slow with a paired decline of $32.9\pct$; launch~10 ran at roughly half speed throughout (paired decline $43.0\pct$). The recorded effective clock stayed ${\approx}1.83$~GHz in both; neither is explained. The reported result uses per-$K$ medians, which move by under 0.03 percentage points if both launches are excluded; all ten paired points are included in the raw data.

\begin{figure}[tbp]\centering
\includegraphics[width=\linewidth]{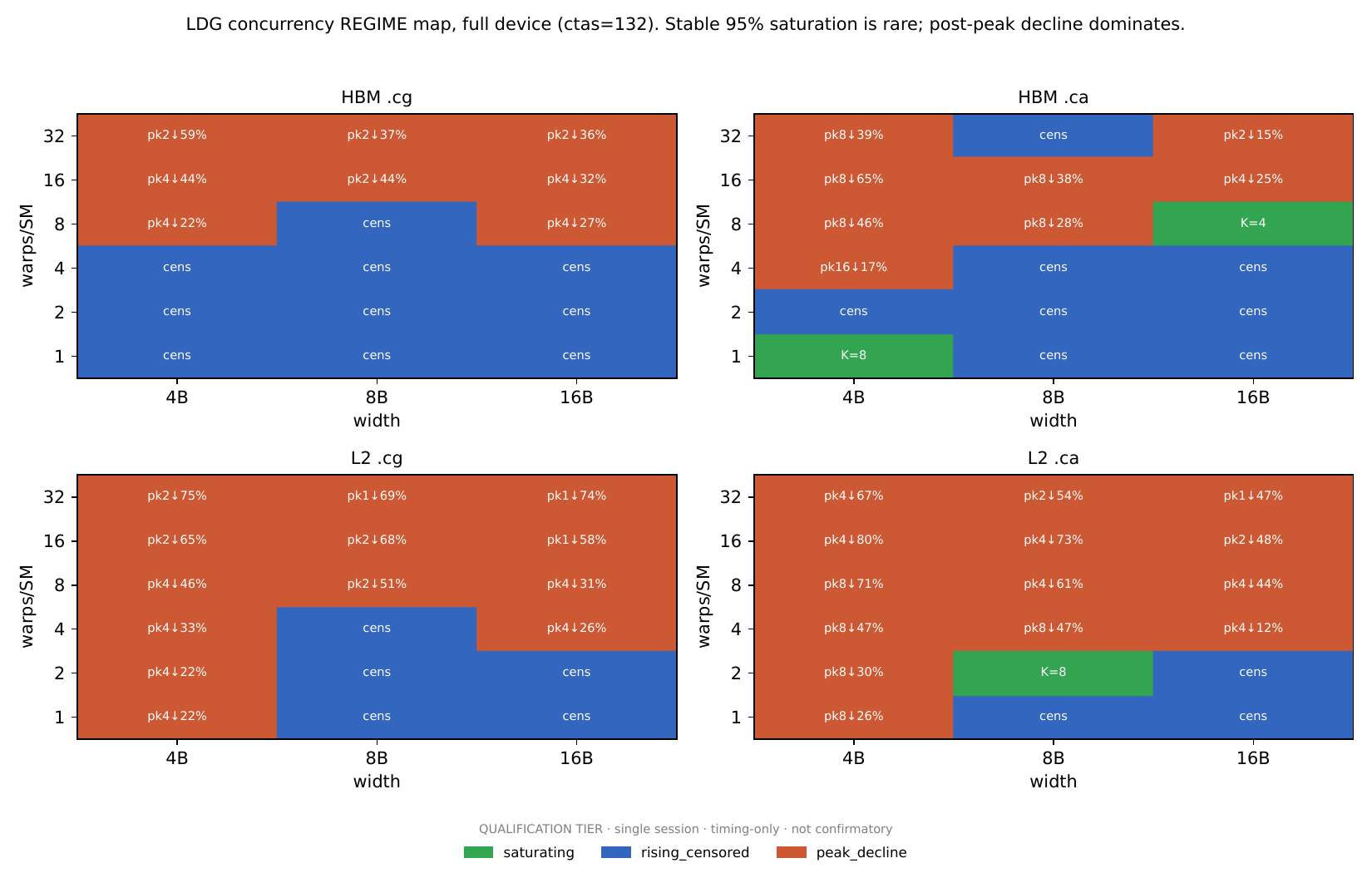}
\caption{Plain-load regime map: each (warps $\times$ width) cell colored saturating, rising-censored, or peak-decline. Stable saturation is rare (8/288); the post-peak decline dominates. Qualification-tier, single die-A session (fixed-iteration sweep).}
\label{fig:regime}
\end{figure}

\paragraph{Prevalence.} Of 288 plain-load response curves on the qualification die, only 8 saturate under the three-cell-tail plateau rule; 179 show the post-peak decline and 101 are still rising at the grid edge (Fig.~\ref{fig:regime}). The decline is more prevalent on the L2-targeted backend than on HBM, and is absent at some narrow-width, low-occupancy cells. This is a configuration-dependent regime in a designed grid, not a universal property of every LDG kernel.

\subsection{\texttt{cp.async} versus a matched plain-load kernel}
\label{sec:cpasync}
\begin{figure}[tbp]\centering
\includegraphics[width=0.82\linewidth]{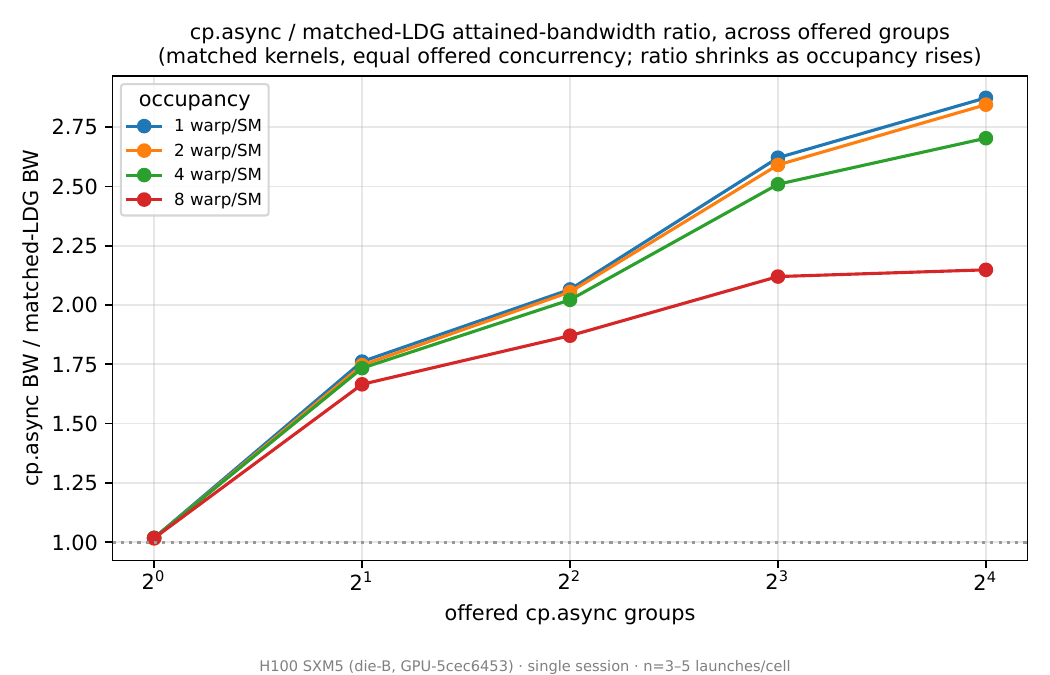}
\caption{\texttt{cp.async} bandwidth relative to a matched plain-load kernel, versus offered groups, per occupancy: $2.1$--$2.9\times$ at 16 groups. (Figure renders the die-B session; the die-C raw data agree.)}
\label{fig:cpasync}
\end{figure}

We compare \texttt{cp.async}, which writes shared memory with no destination register, against a matched \texttt{ld.global}$\to$shared-store kernel at equal offered concurrency (Fig.~\ref{fig:cpasync}). The two agree within about 4\pct{} at a single offered group (die-B $1.02\times$, die-C $1.04\times$); the \texttt{cp.async} advantage grows with offered groups, reaching $2.9\times$ at one warp per SM and $2.1\times$ at eight warps per SM at 16 groups, and shrinks as occupancy rises. Two dies agree (die-B $2.15$--$2.87\times$, die-C $2.10$--$2.81\times$). This is a single-session-per-die, timing-only comparison; we report the measured ratio and do not attribute it to a specific hardware cause.

\subsection{Two co-resident plain-load streams}
\label{sec:twostream}
\begin{figure}[tbp]\centering
\includegraphics[width=0.82\linewidth]{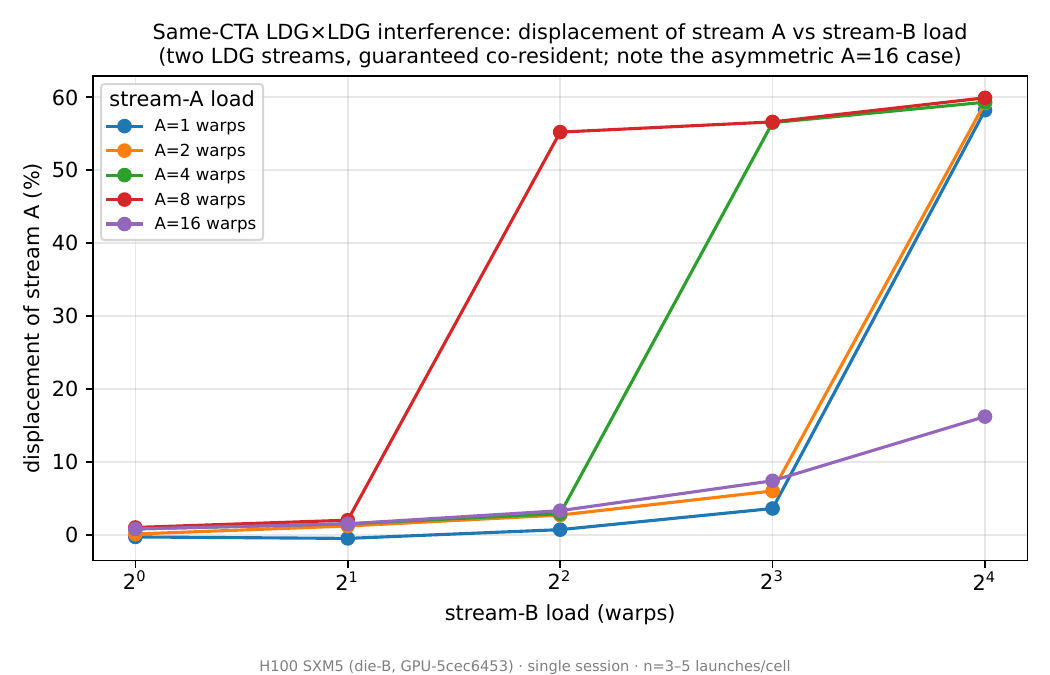}
\caption{Die-B two-co-resident-plain-load-stream series: displacement of stream A versus stream-B load, per probe size. The sampled rise and $A{=}16$ behavior are die-B observations. The companion die-C check differs and is reported in Section~\ref{sec:twostream}; it is not plotted or pooled.}
\label{fig:contention}
\end{figure}

Figure~\ref{fig:contention} reports the die-B single-session same-CTA series. For $A{=}1/2/4/8$ warps, displacement first reaches $58.2/59.2/56.5/55.2\pct$ at $B{=}16/16/8/4$ warps, respectively; $A{=}16$ reaches $16.2\pct$ over the sampled $B$ range. The companion die-C same-CTA grid does not reproduce that shape: at $B{=}16$ its per-$A$ maxima are $12.1/14.6/16.8/20.6/27.8\pct$ for $A{=}1/2/4/8/16$, with $A{=}16$ largest. We therefore report the threshold/asymmetry as a die-B exploratory observation, not as cross-die behavior. This is a same-mechanism (LDG$\times$LDG), same-CTA measurement; it is not a cross-mechanism result, and cross-SM placement is left to future work.

\section{Limitations}
\label{sec:limits}
Only Section~\ref{sec:decline} is fixed-work-controlled, allocation-size-swept, and counter-checked; the \texttt{cp.async} comparison is timing-only across two dies, while the two-stream observation is die-B session-specific and its companion die-C check differs. The counter run audits traffic, not timing. The controlled decline reproduces on two dies (fixed-work) plus a third qualification die in shape; this is not cross-silicon generality, and a full temporal reproducibility bound needs at least four boot sessions. Two of ten primary launches were anomalous (one 0--8\pct{} slow, one at roughly half speed) and remain unexplained. Latency is single-SM, so the near/far L2 structure is not resolved. A separate same-SM loaded-latency probe found the median load latency insensitive to co-resident pressure while the p99 tail rose (from ${\sim}323$ cycles upward); its unloaded-reference gate was unavailable, so we report tail-sensitivity, not a calibrated null. The two-stream measurement is same-CTA only. The mechanism of the decline is not identified: work-volume and simple sweep-order are excluded and a simple allocation-size dependence is disfavored (nominal sweep, no address trace), but aggregate-interleaving and L2-path effects remain open, and naming the mechanism needs stall-reason and throughput-saturation counters we did not collect.

\section{Conclusion and confirmatory work}
\label{sec:concl}
On H100 SXM5 silicon, plain-load bandwidth peaks at a small offered per-thread load and then declines, and this survives a fixed-logical-work control, a reversed-order sweep, and a separately-profiled counter audit on two dies measured with the same instrument; a $40\times$ nominal allocation-size sweep on one of them (die-C, one sweep per allocation $\times$ order cell) leaves the decline essentially unchanged. A preliminary survey adds a two-die \texttt{cp.async}-versus-plain-load ratio and a die-B same-CTA two-stream displacement response; the companion die-C check for the latter differs and is reported rather than pooled. The confirmatory study will add stall-reason and throughput-saturation counters to identify the mechanism, multi-pass statistics on every mechanism, cross-SM placement for the two-stream test, the fixed-work control extended to \texttt{cp.async}, and at least four boot sessions, each behind a written pass/fail gate.

\section{Related work}
\label{sec:related}
Microbenchmark reverse-engineering of GPUs begins with Wong et al. on GT200~\citep{wong2010}. Outstanding-request limits were inferred on Fermi~\citep{nugteren2014} and Kepler~\citep{lashgar2016}; we use the corrected reading of the latter (about 45 table-like entries and roughly 1{,}408 unique outstanding requests, not a 1{,}408-entry table). Volta was dissected in detail by microbenchmark~\citep{jia2018}, its memory system modeled and validated on Titan~V~\citep{khairy2018}, and the memory hierarchy modeled by microbenchmark~\citep{mei2017}, and the Accel-Sim framework validated against such data~\citep{khairy2020}. Recent Hopper-generation work spans Tensor-Core-focused workload characterization~\citep{hanindhito2024}, full-architecture microbenchmark dissection including the asynchronous units~\citep{luo2025}, and compiler support for the asynchronous/TMA paths~\citep{tawa2026}. The closest of these, Luo et al., sweeps memory levels, working-set sizes, and thread/block concurrency and characterizes the TMA and asynchronous-copy paths, but does not vary per-thread offered ILP at fixed logical work or control sweep order, the axes our controlled result requires; Blackwell-generation work applies ILP-sweep microbenchmarks to tensor and memory subsystems~\citep{jarmusch2025}, but on Blackwell rather than Hopper. Relative to these, we add a fixed-work-controlled, two-die plain-load response curve and a die-C $40\times$ nominal allocation-size result that disfavors a simple allocation-size dependence, plus a matched \texttt{cp.async}-versus-LDG comparison and a die-B two-stream exploratory observation with its differing die-C check disclosed. We do not claim priority for measuring any unit.

\section*{Data availability}
The raw measurement logs for all three dies and the analysis scripts that derive every figure and table are available from the authors on reasonable request. The CUDA collection kernels are described in Section~\ref{sec:method}. Statements about three-die primitive agreement are checkable for die-A and die-C; die-B primitives are summarized from its session records.


\end{document}